\documentclass[12pt]{article}
\usepackage[textwidth=6in,top=1.2in,bottom=1.2in]{geometry}
\usepackage{amsthm,amsmath,amssymb,bbm,bm}
\usepackage{natbib}
\usepackage{multirow}
\usepackage[pdftex]{graphicx}
\usepackage{subfigure}
\usepackage{wrapfig}

\usepackage{array}
\usepackage{url}

\usepackage{algorithmic}
\usepackage[linesnumbered, ruled, vlined]{algorithm2e}

\usepackage{mathrsfs}
\usepackage{dsfont}
\usepackage{titling}

\usepackage{relsize}
\usepackage{rotating}
\usepackage{enumitem}
\usepackage{float}
\usepackage[usenames,dvipsnames,svgnames,table]{xcolor}

\usepackage[colorlinks=true,
            linkcolor=red,
            anchorcolor=blue,
            citecolor=blue,
            urlcolor=blue]{hyperref}
\usepackage{setspace}

\newtheorem{remark}{Remark}	
\newtheorem{assumption}{Assumption}[section]

\newtheorem{theorem}{Theorem}[section]
\newtheorem{lemma}{Lemma}

\DeclareMathAlphabet{\pazocal}{OMS}{zplm}{m}{n}

\def\PP{{\mathbbm{P}}}

\def\per{\text{per}}

\makeatletter
\newcommand*{\defeq}{\mathrel{\rlap{%
                     \raisebox{0.3ex}{$\m@th\cdot$}}%
                     \raisebox{-0.3ex}{$\m@th\cdot$}}%
                     =}
\makeatother

\def\pr{\text{pr}}

\newcommand{\cL}{\mathcal{L}}
\newcommand{\cK}{\mathcal{K}}

\newcommand{\yifan}[1]{\textcolor{black}{#1}\xspace}

\begin{document}

\doublespacing

\title{\vspace{-4cm}  {\mbox{Selective machine learning of doubly robust functionals}}}

\author{Yifan Cui, Eric Tchetgen Tchetgen\\Zhejiang University, University of Pennsylvania}

\date{}

\maketitle

\begin{abstract}
While model selection is a well-studied topic in parametric and nonparametric regression or density estimation, selection of possibly high-dimensional nuisance parameters in semiparametric problems is far less developed. In this paper, we propose a selective machine learning framework for making inferences about a finite-dimensional functional defined on a semiparametric model, when the latter admits a doubly robust estimating function and several candidate machine learning algorithms are available for estimating the nuisance parameters. We introduce a new selection criterion aimed at bias reduction in estimating the functional of interest based on a novel definition of pseudo-risk inspired by the double robustness property. Intuitively, the proposed criterion selects a pair of learners with the smallest pseudo-risk, so that the estimated functional is least sensitive to perturbations of a nuisance parameter. We establish an oracle property for a multi-fold cross-validation version of the new selection criterion which states that our empirical criterion performs nearly as well as an oracle with a priori knowledge of the pseudo-risk for each pair of candidate learners. 
Finally, we apply the approach to model selection of a semiparametric estimator of average treatment effect given an ensemble of candidate machine learners to account for confounding in an observational study which we illustrate in simulations and a data application. 
\end{abstract}

\noindent {\bf keywords}
Average Treatment Effect, Doubly Robust, Influence Function,
 Machine Learning, Model Selection

\section{Introduction}\label{sec:intro}

Model selection is a well-studied topic in statistics, econometrics, and machine learning. In fact, methods for model selection and corresponding theory abound in these disciplines, although primarily in settings of parametric and nonparametric regression, where good prediction is the ultimate goal.
 Model selection methods are far less developed in settings where one aims to make inferences about a finite-dimensional, pathwise differentiable functional defined on a semiparametric model. Model selection for the purpose of estimating such a functional may involve selection of an infinite-dimensional parameter, say a nonparametric regression for the purpose of more accurate estimation of the functional in view, which can be considerably more challenging than selecting a regression model strictly for the purpose of prediction. This is because whereas the latter admits a risk, e.g., mean squared error loss,  that can be estimated unbiasedly and therefore can be minimized with small error, the risk of a semiparametric functional will typically not admit an unbiased estimator and therefore may not be minimized without excessive error.  This is an important gap in both model selection and semiparametric theory which this paper aims to address.

 Specifically, we propose a novel approach for model selection of a functional defined on a semiparametric model, in settings where inferences about the targeted functional involve infinite-dimensional nuisance parameters, and the functional of scientific interest admits a doubly robust estimating function.
  Doubly robust inference \citep{robins1994,Rotnitzky1998, Scharfstein1999, robins2001comment,unified, bang2005, cao2009, tan2010, ett2010dr, rotnitzky2012dr, 2015biasreduce, vermeulen2016adaptive,Chernozhukov2018} has received considerable interest in the past few years across multiple disciplines including statistics, epidemiology and econometrics. An estimator is said to be doubly robust if it remains consistent if one of two nuisance parameters needed for estimation is consistent, even if both are not necessarily consistent. The class of functionals that admit doubly robust estimators is quite rich, and includes estimation of pathwise differentiable functionals in missing data problems under missing at random assumptions,  and also in more complex settings where missingness might be not at random.  Several problems in causal inference also admit doubly robust estimating equations, the most prominent of which is the average treatment effect under assumptions that include positivity, consistency and no unmeasured confounding \citep{Scharfstein1999}.
All of these functionals are members of a large class of doubly robust functionals studied by \cite{robins2008HOIF} in a unified theory of first and higher order influence functions.

A well-documented advantage of using doubly robust influence functions is that flexible machine learning or other nonparametric data adaptive methods may generally be used to estimate high-dimensional nuisance parameters. Valid inference may then be obtained about the functional of interest provided that nuisance parameters can be estimated at sufficiently fast rates. Specifically, double robustness ensures that a regular and asymptotically linear estimator of the functional can be obtained if, say both nuisance parameters can be estimated at rates faster than $n^{-1/4}$, although their convergence rates may be considerably slower than the standard parametric rate. Alternatively, it may also be the case that a nuisance parameter can only be estimated at rates slower than $n^{-1/4}$ in which case, by double robustness, sufficiently fast rates of estimation for the other nuisance parameter may still suffice to obtain an estimator of the functional with parametric rate of convergence \citep{robins2008HOIF,robins2017, van2011targeted,van2018targeted,Chernozhukov2018}.  

Several semiparametric approaches have been proposed over the years, which can easily accommodate the use of generic highly adaptive estimators of nuisance parameters to construct, including the use of first order doubly robust influence functions, and more recently higher order influence functions, combined with sample splitting or cross-fitting to construct a one-step estimator \citep{bickel1993efficient,robins2008HOIF,robins2017}; targeted maximum likelihood estimation (TMLE) \citep{van2010collaborative,van2011targeted,van2018targeted,ju2018collaborative,ju2019scalable,ju2019collaborative,Benkeser2020}; and double-debiased machine learning (DDML) \citep{Chernozhukov2018};
A common feature of these existing approaches is their pre-specification of a single nonparametric estimator or machine learning algorithm for estimating each nuisance parameter without regards for the functional ultimately of interest. In contrast, as discussed in Section~\ref{sec:select}, our approach aims to select a machine learning algorithm for each nuisance function with the aim of minimizing bias associated with a possible sub-optimal choice of an estimator for the other nuisance function. We provide theoretical arguments and empirical evidence that such bias awareness in selecting nuisance parameter estimators can provide significant robustness in estimating the functional of interest.

The task of model selection of parametric nuisance models for specific semiparametric doubly robust problems was recently considered by \cite{han2013,chan2013,HAN2014101,han2014jasa,chan2014,duan2017,chen2017,li2020demystifying}, although, their goal differs from ours as they aim to select parametric nuisance models that best approximate each nuisance model, which may generally conflict with selecting the nuisance models that minimize a well-defined risk for the targeted functional. 
In contrast, cross-validated targeted maximum likelihood estimation \citep{zheng2010asymptotic,van2011targeted,van2018targeted} can provide notable improvements on the above methods by allowing the use of an ensemble of highly adaptive estimators of each nuisance parameter, mainly so-called super learner. Still, as indicated above, ensemble learning such as super learner or alternative stacking methods ultimately optimize a well-defined estimation risk associated with each nuisance parameter, and not necessarily a risk directly aimed at estimation of the functional ultimately of interest.

In this paper, we propose an approach for selecting an estimator of nuisance parameters from a large collection of available machine learners, with the aim of constructing a doubly robust estimator of the functional of interest with reduced bias, by minimizing a new quadratic pseudo-risk we introduce. The proposed pseudo-risk is inspired by the double robustness property of the influence function which remains unbiased for the functional in view, provided that one of the nuisance parameters is evaluated at the truth regardless of the degree of mis-specification of the other nuisance parameter. The proposed pseudo-risk is then given by the sum of two maximum squared bias quantities, each capturing the change in the estimated functional, induced by perturbing a nuisance parameter estimator over a collection of candidate learners, say, Lasso, gradient boosting trees, random forest, and neural networks.
As we establish the procedure is guaranteed to recover a consistent estimator for the functional whenever a consistent estimator of either nuisance parameter is available among candidate learners, with corresponding pseudo-risk converging to zero when both estimators are consistent. Our machine learning selector is motivated by \cite{robins2007} who to our knowledge were first to explore potential algorithms to leverage double robustness for the purpose of model selection.
 In order to implement our selective machine learning approach, we propose a cross-validation scheme to estimate the pseudo-risk from observed samples, which we then optimize empirically to select nuisance parameter estimators from available collection of learners. Our main theoretical contribution is to establish an oracle inequality, which states that our empirical selector has pseudo-risk nearly equal to that of an oracle selector with access to the true pseudo-risk. As we also establish that the oracle selector is optimal in the sense that it has bias for the functional of interest that is no larger than the product bias of the estimators of each nuisance parameter with the smallest mean-squared error, this suggests that the empirical selector may be nearly optimal as its pseudo-risk approaches that of the oracle; a possibility we investigate via a simulation study which provides compelling empirical evidence that the selection step indeed provides substantial bias reduction when compared to TMLE and DDML in settings we consider.

The proposed approach is generic, in the sense that it allows the space of candidate machine learning algorithms to be quite large (of order $c^{n^\gamma}$ for any constants $c>0$ and $\gamma<1$), and arbitrary in the sense of including standard parametric, semiparametric, nonparametric, as well as modern machine learning highly adaptive estimators. Importantly, our results are completely agnostic as to whether the collection includes a consistent machine learner for nuisance parameters, in the sense that our procedure will select the pair of candidate machine learning algorithms that optimize our criterion. 
 After completing the selection step, one may wish to report confidence intervals for the functional of interest, anchored at the selected estimator. Three strategies are in fact possible to do so which we discuss in Section~\ref{sec:inference}.

\section{A class of doubly robust functionals} \label{sec:dr}

Suppose we observe $n$ independent and identically distributed samples $\mathcal O \defeq \{O_i,i=1,\ldots,n\}$ from a law $F_0$, belonging to a model $\mathcal M=\{F_\theta: \theta \in \Theta\}$, where $\Theta$ may be infinite-dimensional.
We are interested in inference about a functional $\psi(\theta) \defeq \psi(F_\theta)$ on $\cal M$ for a large class of functionals known to admit a doubly robust first order influence function as defined in \cite{robins2008HOIF}.

\begin{assumption}
 Suppose that  $\theta = \theta_1 \times \theta_2$, where $\times$ denotes Cartesian product, $\theta_1 \in \Theta_1$ governs the marginal law of $X$ which is a $d$-dimensional subset of variables in $O$, and $\theta_2  \in \Theta_2$ governs the conditional distribution of $O|X$.
\label{as1}
\end{assumption}

An influence function is a fundamental object in theory of statistical inference in semiparametric models, that allows one to characterize a wide range of estimators and their efficiency.
The influence function of a regular and asymptotically linear estimator $\widehat \psi$ of $\psi (\theta)$, $\theta \in \cal M$, is a random variable $IF(\theta)\defeq IF(O;\theta)$ which captures the first order asymptotic behavior of $\widehat \psi$, such that ${n}^{1/2}\{\widehat \psi-\psi(\theta)\}=n^{-1/2} \sum_{i=1}^n IF(O_i;\theta) + o_p(1)$.
The set of influence functions of all regular and asymptotically linear estimators of a given functional $\psi(\theta)$ on $\cal M$ is contained in the Hilbert subspace of mean zero random variables $U\defeq u(O;\theta)$ that solve
$d\{\psi(F_t)\}/dt|_{t=0} =E\{US\},$
for all regular parametric submodels $\{F_t:t\in \mathbbm R\}$,
where $S$ is the score function of $f_t$ at $t = 0$ \citep{newey1990semiparametric,bickel1993efficient}.
Once one has identified the influence function of a given estimator, one knows its asymptotic distribution, and can construct corresponding confidence intervals for the target parameter.
We now describe a large class of doubly robust influence functions.
\begin{assumption}
The parameter $\theta_2$ contains components  $p:\mathbbm{R}^d\rightarrow \mathbbm{R}$ and $b:\mathbbm{R}^d\rightarrow \mathbbm{R}$, such that the functional $\psi(\theta)$ of interest has a first order influence function $IF(\theta)= H(p,b)-\psi(\theta)$, where
\begin{align}\label{eq:H}
H(p,b) \defeq b(X)p(X)h_1(O) + b(X)h_2(O) + p(X)h_3(O) + h_4(O),
\end{align}
and $h_1,\ldots,h_4$ are measurable functions of $O$ that do not depend on $\theta$.
\label{as2}
\end{assumption}

\cite{robins2008HOIF} point out that under mild conditions, Assumptions \ref{as1}-\ref{as2} imply the following double robustness property,
\begin{align}\label{eq:dr0}
E_\theta\{H(p^*,b^*)\}-E_\theta\{H(p,b)\}=E[\{p(X)- p^*(X)\}\{b(X)- b^*(X)\}E\{h_1(O)|X\}],
\end{align}
for all $(p^*,b^*)\in \Theta_{2p} \times \Theta_{2b}$, where $\Theta_{2p}$ and $\Theta_{2b}$  are the parameter spaces for the functions $p$ and $b$. In which case $E\{H(p^*,b^*)\}=\psi$ if either $p^*=p$ or $b^*=b$.
Examples of functionals within this class are collected in the Supplementary Material, which include expected product of conditional expectations, expected conditional covariance, missing at random or missing not at random data models, and various functionals that encode average treatment effects, including the population average treatment effect and the effect of treatment on the treated.

The practical implication of double robustness is that an estimator obtained by solving \yifan{$\widehat \psi=\PP_n \{H(\widehat p,\widehat b)\}$} is guaranteed to be consistent provided either but not necessarily both $\widehat p$ is consistent for $p$ or $\widehat b$ is consistent for $b$. Despite this local robustness property, in practice one may be unable to ensure that either model is consistent, and even when using nonparametric models, that the resulting bias is small. For this reason, selection over a class of candidate estimators may be essential to optimize  performance in practical settings.
In this paper, we allow for modern highly adaptive machine learning algorithms to estimate each nuisance function, and consider the increasingly common situation where an analyst must select among a class of machine learning algorithms to estimate each nuisance function.  Hereafter we refer to the task at hand as selective machine learning for inference about a doubly robust functional.

\section{Challenges of selective machine learning for doubly robust inference} \label{sec:prelim}

We describe the proposed selection procedure in estimating average treatment effect as a running example, while the method readily applies to other doubly robust functionals. To simplify the notation, we mainly focus on counterfactual mean $\psi_0=E[Y_1]$, where $Y_1$ is the potential outcome under an intervention that sets treatment to value $1$.
It is well known that under consistency, strong ignorability, and positivity assumptions,
\begin{eqnarray*}
\psi_0  &=&E\left\{ E\left( Y|A=1,X\right)\right\}  \\
&=&E\left\{ \frac{AY}{\pr\left( A=1|X\right) }\right\}  \\
&=&E\left[
\frac{A}{\pr\left( A=1|X\right)}Y
-\left\{ \frac{A}{\pr\left( A=1|X\right)}
E(Y|A=1,X)- E(Y|A=1,X)\right\}
\right].
\end{eqnarray*}

The first representation is known as outcome regression as it depends on the regression of $Y$ on $\left( A,X\right) $; the second is inverse probability weighting with weights depending on the propensity score \citep{10.2307/2335942}; the third representation is known as doubly robust as it relies on outcome regression or propensity score model to be correct but not necessarily both. In fact, the doubly robust representation is based on the efficient influence function of $\psi _{0}$ with $p(X)=1/\pr(A=1|X),$ $b(X)= E(Y|A=1,X)$, $h_1(O)=-A, h_2(O)=1, h_3(O)=AY, h_4(O)=0$ defined in~\eqref{eq:H},
which will be used as our estimating equation for $\psi_0$ for the proposed selection procedure.

For simplicity, in a slight abuse of notation, we denote $p(X)=\pr(A=1|X)$ and $b(X) = E(Y|A=1,X)$, and $p$ here technically equals to the reciprocal of $p$ defined in $H(p,b)$ in Section~\ref{sec:dr}.
 Suppose that we have machine learners $p_k\left(X\right),  k \in \mathcal K \defeq \{1,\ldots,K\}$ for the propensity score and $b_l \left(X\right), l \in \mathcal L \defeq \{1,\ldots,L\}$ for the outcome model, respectively.
In order to describe the inherent challenges of performing selection of machine learners for $\psi_0$,
consider the sample splitting scheme whereby a random half of the sample, i.e., training sample,  is used to construct  $\widehat p_k(X)$ and $\widehat b_l(X)$, while the other half is used to obtain the doubly robust estimator \yifan{by $\widehat \psi_{k,l}=\PP^1_n [H(\widehat p_k,\widehat b_l)]$},
where $\PP^1_n$ denotes the empirical measure with respect to the validation sample.

Consider the goal of selecting a pair of models $(k, l)$ that minimizes the mean squared error $\PP^1\{(\widehat \psi_{k, l}-\psi_0)^2\}$ = bias$^2(\widehat \psi_{k,l})$ + variance$(\widehat \psi_{k,l})$,
where $\PP^1$ denotes the expectation with respect to the validation sample, and bias$^2(\widehat \psi_{k,l})$ is given by
  \begin{align}
\text{bias}^2(\widehat \psi_{k,l}) & = \left[\PP^1\left(
\frac{A}{\widehat p_k \left(X\right) }Y
-\left\{ \frac{A}{\widehat p_k \left( X\right) }
\widehat b_l(X)- \widehat b_l(X)\right\}\right)-\psi_0 \right]^2 \nonumber \\
&=\left[\PP^1 \left\{\frac{p(X)}{\widehat p_k(X)}-1 \right\} \left\{b(X)-\widehat b_l(X)\right\}\right]^2.
 \label{eq:risk2}
 \end{align}

As we expect the variance term to be of order $1/n$ conditional on training sample, we may focus primarily on minimizing the squared bias.
\yifan{To the best of our knowledge, no unbiased estimator of $\text{bias}(\widehat \psi_{k,l})$ exists without an additional assumption, otherwise it would be possible to obtain an estimator of the functional of interest with influence function that, contrary to fact, does not depend on a nuisance parameter \citep{ibragimov2013statistical}. Therefore, minimizing $\text{bias}^2(\widehat \psi_{k,l})$ will generally not be possible without incurring additional uncertainty, without an additional assumption about the underlying data generating process.}
In the next section, we propose an alternative criterion for selecting an estimator with a certain optimality condition that is nearly attainable empirically without further restrictions about the underlying data generating process.

\begin{remark}
Recall that consistent estimators of the propensity score and outcome regression are not necessarily contained as candidates for selection, so the minimal squared bias may not necessarily converge to zero asymptotically; nevertheless, it will do so when at least one nuisance parameter is estimated consistently. Furthermore, as we formally establish in Sections~\ref{sec:theory1} and \ref{sec:theory2} and illustrate in our simulations,
when a library of flexible machine learning estimators is used to estimate nuisance parameters, the approach proposed in the next section behaves nearly as well as an oracle that selects the estimator with the smallest average squared bias, which vanishes at least as fast as any given choice of machine learners. This is quite remarkable as the proposed approach avoids directly estimating the squared bias.
\end{remark}

\section{Selective machine learning via mixed-minimax cross-validation} \label{sec:select}

\subsection{A mixed-minimax criterion for selective machine learning} \label{sec:selection}
In this section, we consider a selection criterion which avoids estimating and directly minimizing Equation \eqref{eq:risk2}.
We begin by describing the population version of our minimax criterion, i.e., we focus on $p_k \left( X\right)$ and $b_l(X)$, the asymptotic limits of $\widehat p_k\left(X\right)$ and $\widehat b_l(X)$. 
The propensity score and outcome learners could be parametric, semiparametric or nonparametric.
We will introduce the cross-validated estimator in Section~\ref{sec:cf}.
For each pair of candidate learners $(k, l)$, we have
\begin{align}
\psi _{k,l} = \PP^1\left[\frac{A}{p_{k}\left(X\right) }Y
-\left\{ \frac{A}{p_{k}\left(X\right) }
b_{l}(X) - b_{l}(X)\right\}\right],
\label{eq:if2}
\end{align}
where $\psi_{k,l}$ is the probability limit of  
$\PP^1_n \{H(\widehat p_k,\widehat b_l)\}$.

\yifan{If} $p_k = p$ the functionals in (4) agree for all $l$; this suggests
using some measure of spread of the estimators 
$\left\{\widehat \psi_{k,l}  ,l\in \cL\right\}$ as a basis for selecting between the machine learners for $p$; and likewise for
choosing the estimator of $b$. 
Motivated by this observation, we define the following perturbation of a fixed index pair $(k_0,l_0)$,
\begin{align}
\per(k,l; k_0,l_0) \defeq (\psi_{k,l}- \psi_{k_0,l_0})^2.
\label{definition:per}
\end{align}
Note that one has 
\begin{align*}
\{\max_{l_1,l_2\in \cL} \text{per}(k,l_1,k,l_2)\}^{1/2} = &
\max_{l\in \cL} \psi_{k,l}- \min_{l\in \cL} \psi_{k,l}
\defeq  \text{range}(\psi_{k,\cdot}),
\end{align*}
and likewise
\begin{align*}
\{\max_{k_1,k_2\in \cK} \text{per}(k_1,l,k_2,l)\}^{1/2} = &
\max_{k\in \cK} \psi_{k,l}- \min_{k\in \cK} \psi_{k,l}
\defeq  \text{range}(\psi_{\cdot,l}).
\end{align*}
Therefore, we consider the following population pseudo-risk,
\begin{align*}
B_{k_0,l_0}=\max_{l_1,l_2 \in \mathcal L} \text{per}(k_0,l_1; k_0,l_2) + \max_{k_1,k_2 \in \mathcal K} \text{per}(k_1,l_0; k_2,l_0).
\end{align*}
We call this a pseudo-risk because unlike a standard definition of risk (e.g., mean squared error) which is typically defined in terms of the data generating mechanism and a given candidate machine learner, the proposed definition is in terms of all candidate machine learners.
For instance, for $K=2,L=2$ and $K=3,L=3$, 
\begin{align*}
B_{1,1} = \text{per}(1, 1; 1, 2) + \text{per}(1, 1; 2, 1),
\end{align*}
and 
\begin{align*}
B_{1,1} = & \max\{\text{per}(1, 1; 1, 2), \text{per}(1, 1; 1, 3), \text{per}(1, 2; 1, 3)\}
\\ & + \max\{\text{per}(1, 1; 2, 1), \text{per}(1, 1; 3, 1), \text{per}(2, 1; 3, 1)\},
\end{align*}
respectively. 

Evaluating the above perturbation for each pair $(k_0,l_0)$ gives $K \times L$ pseudo-risk values $B_{k_0,l_0}, k_0\in \mathcal K; l_0 \in \mathcal L$. Finally, we define
\begin{align*}
\arg\min_{(k,l)} B_{k,l},
\end{align*}
as the population version of selected learners, respectively. The mixed-minimax criterion admits a doubly robust property, i.e.,  $\psi_{\arg\min_{(k,l)} B_{k,l}} = \psi_0$ if either selected nuisance learner is consistent. It can be seen easily from the following expression, 
\begin{align*}
\arg\min_{k\in \cK,l\in \cL} B_{k,l} = (
\arg \min_{k\in \cK} \{\text{range}(\psi_{k,\cdot})\}, \arg \min_{l\in \cL} \{\text{range}(\psi_{\cdot,l})\}).
\end{align*}
The selector based on the pseudo-risk selects the learner for $p$ that minimizes some measure of spread
for the distinct estimators that use a fixed $p_k$ and likewise for the selectors of $b$. In particular, if a learner $\widehat p_{k^*}$ is consistent for $p$ then $k^*\in \arg\min_k \{\text{range}(\psi_{k,\cdot})\}$
 and in large samples with probability going to 1, the
procedure chooses the right learner $k^*$. This fact is reflected in the result established in Lemma~\ref{lemma:rate} in Section~\ref{sec:theory1}.


\subsection{Multi-fold cross-validated estimators} \label{sec:cf}
Following \cite{vaart2006oracle}, we avoid overfitting by considering a multi-fold cross-validation scheme which repeatedly splits the data $\mathcal O$ into two subsamples: a training set $\mathcal O^{0s}$ and a validation set $\mathcal O^{1s}$, where $s$ refers to the $s$-th split. The splits may be either deterministic or random without loss of generality. In the following, we consider random splits, whereby we let $T^s = (T_1^s,\ldots,T_n^s)\in \{0,1\}^n$ denote a random vector independent of $\{O_1,\ldots, O_n\}$. If $T_i^s=0$, $O_i$ belongs to the $s$-th training sample $\mathcal O^{0s}$; otherwise it belongs to the $s$-th validation sample $\mathcal O^{1s}$, $s=1, \ldots,S$.
For each $s$ and $(k,l)$, our construction uses the training samples to construct estimators $\widehat p_k^s(X)$ and $\widehat b_l^s(X)$.
The validation sample is then used to estimate the perturbation defined in Equation~\eqref{definition:per},
\begin{align*}
\widehat {\text{per}}(k,l; k_0,l_0) =& \frac{1}{S}\sum_{s=1}^S \left\{  (\widehat \psi_{k,l}^s  - \widehat \psi_{k_0,l_0}^s )
\right\}^{2},
\end{align*}
where \yifan{$\widehat \psi^s_{k,l}=\PP^1_s [H(\widehat p^s_k,\widehat b^s_l)]$},
$$\PP^j_s= \frac{1}{\#\{1\leq i\leq n:T_i^s=j\}}\sum_{i:T_i^s=j} \delta_{X_i}, \quad j=0,1,$$
and $\delta_X$ is the Dirac measure.
We then select the empirical minimizer of $$\widehat B_{k_0,l_0} =\max_{l_1,l_2 \in \mathcal L} \widehat{\text{per}}(k_0,l_1; k_0,l_2) + \max_{k_1,k_2 \in \mathcal K} \widehat{\text{per}}(k_1,l_0; k_2,l_0),$$ among all candidate pairs $(k_0,l_0)$ as our selected learners, denoted by $(\widehat k^*,\widehat l^*)$.
The final estimator is $$\widehat \psi_{\widehat k^*,\widehat l^*}=\frac{1}{S}\sum_{s=1}^S \widehat \psi^s_{\widehat k^*,\widehat l^*}.$$
\yifan{The practical choice of $S$ should be based on empirical performance. In simulations studies, we have found that $S=2$ performed quite well for sample sizes 250-2000.}
We provide a high-level Algorithm~\ref{alg:cf} for the proposed selection procedure in Appendix.

\section{Theoretical results\label{sec:theory}}

\subsection{Optimality of the oracle selector\label{sec:theory1}}
In this section, we establish certain optimality property of the oracle pseudo-risk selector defined by 
\yifan{\begin{align}
 (k^{*},l^{*}) = \arg \min_{(k_0,l_0)}
\left[ \max_{l_1,l_2 \in \mathcal L} \frac{1}{S}\sum_{s=1}^S \left\{  (\widehat \psi_{k_0,l_1}^{s,or}  - \widehat \psi_{k_0,l_2}^{s,or} )
\right\}^{2} + \max_{k_1,k_2 \in \mathcal L} \frac{1}{S}\sum_{s=1}^S \left\{ (\widehat \psi_{k_1,l_0}^{s,or}  - \widehat \psi_{k_2,l_0}^{s,or} )
\right\}^{2} \right],
  \label{eq:defstar}
 \end{align}
 where $\widehat \psi^{s,or}_{k,l}=\PP^1 [H(\widehat p^s_k,\widehat b^s_l)]$.
 Without loss of generality, we consider $S=1$ in this subsection.}
As we will later show by establishing excess risk bound relating empirical selector to its oracle counterpart, this
optimality result implies near optimal behavior of the corresponding
empirical (cross-validated) selector. In this vein, focusing on a functional
of the doubly robust class with property \eqref{eq:dr0}, consider the collections of learners for
$p$ and $b$ obtained from an independent sample of size $n$:
\begin{eqnarray*}
\mathcal{C}_{p} = \left\{ \widehat{p}_{1},\ldots,\widehat{p}_{K}\right\};~~
\mathcal{C}_{b} = \left\{ \widehat{b}_{1},\ldots,\widehat{b}_{L}\right\}.
\end{eqnarray*}%
For the purpose of inference, in the following, our analysis is conditional on $\mathcal{C}_{p}$ and $\mathcal{C}_{b}$. Suppose further that these learners satisfy the following assumptions.

\begin{assumption}\label{asm:1} Given any $\epsilon>0$, there exist constants $C_p, C_b>1$ and sufficiently large $n_0$ such that for $n>n_0$,
\begin{align*}
\frac{1}{C_{p}}\nu _{k}\leq E\{h_1(O)|X=x\}^{1/2}\vert \widehat{p}_{k}(x)-p\left( x\right)
\vert \leq C_{p}\nu _{k},~~ k\in \mathcal K,\\
\frac{1}{C_{b}}\omega _{l}\leq E\{h_1(O)|X=x\}^{1/2}\vert \widehat{b}_{l}(x)-b\left(
x\right) \vert \leq C_{b}\omega _{l},~~ l\in \mathcal L,
\end{align*}%
for any $x$ with probability larger than $1-\epsilon$, where $\nu _{k}$ and $\omega_{l}$ depend on $n$.
\end{assumption}

Let $\nu _{\max } =\max_{k}\left\{ \nu _{k}:k \in \mathcal K \right\}$, $\omega _{\max } =\max_{l}\left\{ \omega _{l}:l \in \mathcal L\right\}$, $\nu _{\min }=\min_{k}\left\{ \nu _{k}:k \in \mathcal K\right\}$, and $\omega
_{\min }=\min_{l}\left\{ \omega _{l}:l \in \mathcal L\right\}$.
\begin{assumption}\label{asm:2}
We assume that
$\lim_{n\rightarrow \infty }v_{\max } < \infty, ~~
\lim_{n\rightarrow \infty }\omega _{\max } < \infty.$
\end{assumption}

\begin{assumption}\label{asm:3}
Suppose $$\{p(X)- \widehat p_k(X)\}\{b(X)- \widehat b_l(X)\}E\{h_1(O)|X\},$$ is continuous with respect to $X$ for $k\in \mathcal K; l\in \mathcal L$. Furthermore, suppose that the support of $X$ is closed and bounded.
\end{assumption}

Assumption~\ref{asm:1} essentially states that the bias of $\widehat p_k$  and $\widehat b_l$ is eventually exactly of order $v_k$ and $w_l$ with large probability.  Note that $\widehat p_k$  and $\widehat b_l$ may not necessarily be consistent, i.e., $v_k$ and $w_l$ may converge to a positive constant.
Assumption~\ref{asm:2} guarantees the bias of each learner does not diverge. 
\yifan{Let $\widehat \psi^{or}_{k^*,l^*} =\PP^1 \{H(\widehat{p}_{k^{*}},\widehat{b}_{l^{*}})\}$, where $H(\cdot,\cdot)$ is defined in Equation~\eqref{eq:H}, and $(k^{*},l^{*})$ is defined in Equation~\eqref{eq:defstar} with setting $S=1$.}
We have the following lemma.
\begin{lemma} \label{lemma:rate} Under Assumptions~\ref{asm:1}-\ref{asm:3}, we have that the bias of the mixed-minimax oracle selector is of the order of:
\yifan{\begin{equation*}
\left \vert \widehat \psi^{or}_{k^*,l^*}-\psi_0
\right \vert =O_{P}\left( \nu _{\min }\omega _{\min }\right).
\end{equation*}}
\end{lemma}

This lemma implies that the mixed-minimax
selector can leverage doubly robust property to select the best estimator, so that in the
above scenario, its rate of estimation would be $\nu _{\min }\omega _{\min
}$.

\subsection{Excess risk bound of the proposed selector\label{sec:theory2}}

Define $U^s_{(k,l)}(k_0,l_0)\defeq H(\widehat p^s_k,\widehat b^s_l)-H(\widehat p^s_{k_0},\widehat b^s_{l_0})$.
Based on our mixed-minimax selection criterion, $(\widehat k^{*},\widehat l^{*})$ and $(k^{*},l^{*})$ are equivalently expressed as
$$(\widehat k^{*},\widehat l^{*}) = \arg\min_{(k_0,l_0)} \left(\max_{l_1,l_2\in \cL} \frac{1}{S}\sum_{s=1}^{S} [ \PP_s^1 \{U^s_{(k_0,l_1)}(k_0,l_2) \}]^2 + \max_{k_1,k_2\in \cK} \frac{1}{S}\sum_{s=1}^{S} [ \PP_s^1 \{U^s_{(k_1,l_0)}(k_2,l_0) \}]^2 \right),$$
$$(k^{*},l^{*}) = \arg\min_{(k_0,l_0)} \left(\max_{l_1,l_2\in \cL} \frac{1}{S}\sum_{s=1}^{S} [ \PP^1 \{U^s_{(k_0,l_1)}(k_0,l_2) \}]^2 + \max_{k_1,k_2\in \cK} \frac{1}{S}\sum_{s=1}^{S} [ \PP^1 \{U^s_{(k_1,l_0)}(k_2,l_0) \}]^2 \right).$$

Next, we derive a risk bound for empirically selected learners $(\widehat k^{*},\widehat l^{*})$ which states that its risk is not much bigger than the risk provided by the oracle selected learners $(k^{*},l^{*})$. For this purpose, \yifan{we make the following boundedness assumption.}
\begin{assumption}\label{asm:positivity}
\yifan{(1) $|p(X)|$ and $|\widehat p_{k}(X)|$ are bounded almost surely, where $k \in \mathcal K$. (2) $|b_l(X)|$ and $|\widehat b_{l}(X)|$ are bounded almost surely, where $l \in \mathcal L$. (3) $|h_1(O)|$, $|h_2(O)|$, $|h_3(O)|$ and $|h_4(O)|$ are bounded almost surely.}
\end{assumption}

\begin{theorem}
Suppose Assumption~\ref{asm:positivity} holds, then we have that
\yifan{
\begin{align*}
&  \frac{1}{S}\sum_{s=1}^{S} \PP^0[\PP^1 \{U^s_{(\widehat k^{*},\widetilde l_1^{*})}(\widehat k^{*},\widetilde l_2^{*})\}]^2 \\
\leq & \frac{1+2\delta}{S}\sum_{s=1}^{S} \PP^0[\PP^1 \{U^s_{(k^{*},\bar l_1^{*})}(k^{*},\bar l_2^{*})\}]^2  +  \frac{C_1(1+\delta)\log(1+KL^2)}{n^{1/q}} \left(\frac{1+\delta}{\delta}\right)^{(2-q)/q},\\
&  \frac{1}{S}\sum_{s=1}^{S} \PP^0[\PP^1 \{U^s_{(\widetilde k_1^{*},\widehat l^{*})}(\widetilde k_2^{*},\widehat l^{*})\}]^2\\
\leq & \frac{1+2\delta}{S}\sum_{s=1}^{S} \PP^0[\PP^1 \{U^s_{(\bar k_1^{*},l^{*})}(\bar k_2^{*},l^{*})\}]^2 + \frac{C_2(1+\delta)\log(1+LK^2)}{n^{1/q}} \left(\frac{1+\delta}{\delta}\right)^{(2-q)/q},
\end{align*}
}
for any $\delta>0$, \yifan{any} $1\leq q\leq 2$, and some constants $C_1,C_2$, where
$$(\widetilde k_1^{*},\widetilde k_2^{*})=\arg\max_{k_1,k_2} \frac{1}{S} \sum_{s=1}^S [\PP^1_s\{U^s_{(k_1,\widehat l^{*})}(k_2,\widehat l^{*})\}]^2,(\widetilde l_1^{*},\widetilde l_2^{*})=\arg\max_{l_1,l_2} \frac{1}{S} \sum_{s=1}^S [\PP^1_s\{U^s_{(\widehat k^{*},l_1)}(\widehat k^{*},l_2)\}]^2,$$ $$(\bar k_1^{*},\bar k_2^{*})=\arg\max_{k_1,k_2} \frac{1}{S} \sum_{s=1}^S [\PP^1\{U^s_{(k_1, l^{*})}(k_2, l^{*})\}]^2,(\bar l_1^{*},\bar l_2^{*})=\arg\max_{l_1,l_2} \frac{1}{S} \sum_{s=1}^S [\PP^1\{U^s_{( k^{*},l_1)}(k^{*},l_2)\}]^2,$$
and $\PP^0$ denotes the expectation with respect to training data.
\label{thm:2}
\end{theorem}

The proof of this result is based on a finite sample inequality for the excess pseudo-risk of our selector compared to the oracle's,
which requires an extension of Lemma 8.1 of \cite{vaart2006oracle} to handle second order U-statistics.
We obtained such an extension by making use of a powerful exponential inequality for the tail probability of the maximum of a large number of second order degenerate U-statistics derived by \cite{10.1007/978-1-4612-1358-1_2}. Note that Theorem~\ref{thm:2} generalizes to the doubly robust functionals in the class of \cite{rotnitzky2019mix}, with $U^s_{(k,l)}(k_0,l_0)$ replaced by $IF(\widehat p^s_{k},\widehat b^s_{l} ,\widehat \psi^s_{k_0,l_0})-IF(\widehat p^s_{k_0},\widehat b^s_{l_0}, \widehat \psi^s_{k_0,l_0})$ (see Algorithm~2 in the Supplementary Material for details).

\yifan{The bound given in Theorem~\ref{thm:2} is valid for any $\delta>0$, such that the error incurred by empirical risk is of order $n^{-1}$ for any fixed $\delta$ if $q=1$, suggesting in this case that our cross-validated selector performs nearly as well as an oracle selector with access to the true pseudo-risk.
Theorem~\ref{thm:2} is of interest in a nonparametric/machine learning setting where pseudo-risk can be of order substantially larger than $O(n^{-1})$ in which case the error made in selecting optimal learners is negligible relative to its risk.
Furthermore, the derived excess risk bound holds for as many as $c^{n^\gamma}$ models for any $\gamma<1$ and $c>0$.}

\yifan{
\subsection{Double robustness of the proposed selector\label{sec:theory3}}}

\yifan{
In this section, by leveraging the results developed in Sections~\ref{sec:theory1} and \ref{sec:theory2}, we establish that under certain conditions, the selective machine learning estimator is in fact doubly robust. 
In this vein, suppose that each candidate learner $p_k,b_l$ has a distinct probability limit for every $x$, where $k\in \cK,l\in \cL$; this implies that at most one estimator of $b$ and $p$ is consistent, respectively. In addition, suppose that either of the following two assumptions holds.
}

\yifan{
\begin{assumption}\label{asm:mono1}
Suppose that candidate learners are monotone, that is, $\widehat p_{(1)}(x) < \widehat p_{(2)}(x) < \ldots  < \widehat p_{(K)}(x)$ and $\widehat b_{(1)}(x) < \widehat b_{(2)}(x) \ldots < \widehat b_{(L)}(x)$ for every $x$.
\end{assumption}
\begin{assumption}\label{asm:mono2}
Suppose that $\widehat p_{\widehat k_2}(x) < p(x)<\widehat p_{\widehat k_1}(x)$ (or $\widehat p_{\widehat k_1}(x) < p(x)<\widehat p_{\widehat k_2}(x)$) and $\widehat b_{\widehat l_2}(x) < b(x)< \widehat b_{\widehat l_1}(x)$ (or $\widehat b_{\widehat l_1}(x) < b(x)< \widehat b_{\widehat l_2}(x)$) for every $x$, where 
 \begin{align}
(\widehat k_1,\widehat k_2)=& \arg\max_{k_1,k_2 \in \cK}|E[
\{\widehat p_{k_2}(X) - \widehat p_{k_1}(X)\}\{b(X)- \widehat b_{\widehat l^*}(X)\}E\{h_1(O)|X\}]|, \label{eq:addassm1}\\
(\widehat l_1,\widehat l_2)=& \arg\max_{l_1,l_2 \in \cL}|E[
\{p(X) - \widehat p_{\widehat k^*}(X)\}\{\widehat b_{l_2}(X)- \widehat b_{l_1}(X)\}E\{h_1(O)|X\}]|. \label{eq:addassm2}
\end{align}
\end{assumption}
 Assumption~\ref{asm:mono1} restricts the candidate learners to certain classes.  For example, for sieve or series-type  regression including Lasso-type penalty, it essentially places a restriction on slopes and intercepts.
 We note that it is only a sufficient condition for our results. 
  Assumption~\ref{asm:mono2} supposes that the selected learners $\widehat p_{\widehat k_1}(x)$ and $\widehat p_{\widehat k_2}(x)$ in \eqref{eq:addassm1} and $\widehat b_{\widehat l_1}(x)$ and $\widehat b_{\widehat l_2}(x)$  in \eqref{eq:addassm2} are uniformly bounded above and below by the true nuisance function $p(x)$ and $b(x)$, respectively; this would be the case if selected estimators that maximize the risk inside minimax objective function, in fact bracket the true function.
  Furthermore, we also make the following assumption on $h_1(O)$.
\begin{assumption}\label{asm:h1}
We assume that $h_1(O)$ does not depend on $X$ and suppose $E\{h_1(O)\}\neq 0$.
\end{assumption}
Assumption~\ref{asm:h1} holds for the running counterfactual mean example as well as all five examples provided in the Supplementary Material. We then have the following theorem. 
}

\yifan{
\begin{theorem}\label{thm:3}
Suppose that either class of candidate machine learners $\mathcal C_p$ or $\mathcal C_b$ includes at least one consistent estimator.
Under Assumptions~\ref{asm:1}-\ref{asm:positivity}, \ref{asm:h1} and suppose either Assumption~\ref{asm:mono1} or \ref{asm:mono2} holds, we have that
 $\widehat \psi_{\widehat k^*,\widehat l^*}\xrightarrow{p}
\psi$.
\end{theorem}
}\yifan{
The above theorem verifies that our selector is doubly robust in the sense that if either class of candidate machine learners includes a consistent estimator, the proposed selector is consistent.
}

\section{Challenges of valid inference for selective machine learning} \label{sec:inference}

It is challenging to propose a valid mode of inference for the proposed selective machine learning framework because of dependence across cross-validation samples. 
We discuss three possible strategies to implement confidence intervals for practitioners:

(i) One may report conventional Wald confidence intervals using the influence function for the functional, evaluated at selected learners for the nuisance parameters. This may be the simplest strategy which appropriately accounts for uncertainty in the estimated nuisance parameters, however, as it is completely blind to the model selection step, it may not yield uniformly valid confidence intervals for the functional of interest.

\yifan{Specifically, 
 the standard error of the proposed estimator is computed by the covariance matrix of the influence functions evaluated at the selected $\widehat \psi_{\widehat k^*,\widehat l^*}$ and selected nuisance function learners $(\widehat p_{\widehat k^*}, \widehat b_{\widehat l^*})$ from cross-validation. 
Taking a two-fold cross-validation $S=2$ as an example, the variance of the proposed estimator $(\widehat \psi^1_{\widehat k^*,\widehat l^*} + \widehat \psi^2_{\widehat k^*,\widehat l^*})/2$ is computed by the variance of $\widehat \psi^1_{\widehat k^*,\widehat l^*}$, the variance of $\widehat \psi^2_{\widehat k^*,\widehat l^*}$, and the covariance of $\widehat \psi^1_{\widehat k^*,\widehat l^*}$ and $\widehat \psi^2_{\widehat k^*,\widehat l^*}$.
The variance of $\widehat \psi^1_{\widehat k^*,\widehat l^*}$ and $\widehat \psi^2_{\widehat k^*,\widehat l^*}$ can be computed by the variance of
\begin{align*}
\widehat{IF}^i = \widehat b_{\widehat l^*}(X)\widehat p_{\widehat k^*}(X)h_1(O) +  \widehat b_{\widehat l^*}(X)h_2(O) + \widehat p_{\widehat k^*}(X)h_3(O) + h_4(O)-\widehat \psi^i_{\widehat k^*,\widehat l^*},
\end{align*}
where $i=1,2$, respectively. 
The covariance of $\widehat \psi^1_{\widehat k^*,\widehat l^*}$ and $\widehat \psi^2_{\widehat k^*,\widehat l^*}$ can be computed by 
\begin{align*}
n_0 \frac{\text{var}(\widehat \psi^1_{\widehat k^*,\widehat l^*})n_1+\text{var}(\widehat \psi^2_{\widehat k^*,\widehat l^*})n_2}{2n_1 n_2},
\end{align*}
where $n_1=\#\{1\leq i\leq n:T_i^1=1\},n_2=\#\{1\leq i\leq n:T_i^2=1\}$ are validation sample sizes and $n_0$ is the number of samples fall into both validation sets.
}

(ii) An alternative approach one could consider is to construct a cross-fitted estimator whereby one part of the data is used to perform the model selection step and the other is used for estimating the functional using selected nuisance learners; the procedure would then be repeated upon swapping the roles of the two samples and an estimator and corresponding confidence intervals could be obtained by a weighted average of the two estimators and their corresponding influence functions to produce  uniformly valid confidence interval under standard conditions for cross-fitting; however, further splitting of the sample is likely to induce more variability in the learner selection step as the sample used is effectively a fraction of the original sample.

(iii) Finally, one  may construct a smooth approximation of the proposed selector which allows for valid post-selection inference. This essentially corresponds to a smooth version of stacking that adds stability to the chosen ensemble of learners, while at the same time concentrating most weight on the best performing learners.  Recent theoretical advances in machine learning \citep{austern2020asymptotics} suggest a single split version of our selector may justify the use of the bootstrap to evaluate uncertainty.

Of these three possible solutions, (i) is explored both in simulations and the data application, while (iii) is described in detail in Section~S4 of the Supplementary Material; more formal comparisons of these three approaches are beyond the scope of the current paper which is primarily focused on the potential for bias reduction provided by the proposed model selection approach.


\section{Numerical experiments}\label{sec:simu3}
In this section, we report simulation results allowing for various machine learners as candidate estimators of  nuisance parameters. We considered the following machine learning methods.
For the propensity score model:
1. Logistic regression with $l_1$ regularization \citep{lasso,friedman2010regularization}; 2. Classification random forests \citep{Breiman2001,ranger}; 3. Gradient boosting trees \citep{friedman2001greedy,gbm2019}; 4. Super learner \citep{van2011targeted}.
For the outcome model: 1. Lasso; 2. Regression random forests; 3. Gradient boosting trees; 4. Super learner.
Data were generated from
\begin{eqnarray*}
\text{logit} \{\pr \left( A=1|X\right)\} &=&(1,-1,1,-1,1)^Tf(X),\\
E\left( Y|A,X\right) &=& 2(1 + \mathbbm{1}^Tf(X)+ \mathbbm{1}^Tf(X) A+ A),
\end{eqnarray*}
where $X$ followed a uniform distribution on $(0, 1)$, the outcome error term followed standard normal distribution, and we considered the following five-variate functional forms \citep{zhao2017selective} for the first scenario,
\begin{align*}
f(x) &= \Big(\frac{1}{1+\exp\{-20(x_1-0.5)\}}, \ldots, \frac{1}{1+\exp\{-20(x_5-0.5)\}}\Big)^T,
\end{align*}
and
\begin{align*}
f(x) &= \Big(x_1^2,x_2^2,\ldots,x_5^2\Big)^T,
\end{align*}
\yifan{for the second scenario.}
Implemented candidate estimators used default tuning parameters in \cite{superlearner}.
The proposed selection procedure was implemented with two-fold cross-validation.
Each simulation was repeated 500 times.

\yifan{
We compared the proposed estimators with four DDML estimators \citep{Chernozhukov2018} and four cross-validated TMLE (CV-TMLE)  \citep{van2011targeted}
using Lasso, random forests, gradient boosting trees, and super learner to estimate nuisance parameters respectively, where the super learner uses a library consisting of Lasso, random forest, and gradient boosting trees.
We have deliberately restricted our comparisons to CVTMLE as appears to be the most widely used TMLE estimator and did not consider other TMLE variants such a C-TMLE and super-efficient TMLE, though we briefly mention them as alternative TMLE estimators one might consider.
Each DDML was estimated by cross-fitting \citep{Chernozhukov2018}, i.e., 1) using training data (random half of sample) to estimate nuisance parameters and validation data to obtain $\widehat \psi_1$; 2) swapping the role of training and validation dataset to obtain $\widehat \psi_2$; 3) computing the estimator as the average $\widehat \psi_{\text{CF}} = (\widehat \psi_1 + \widehat \psi_2)/2$. 
Each CV-TMLE was estimated by cross-fitting and the tuning parameters follow the default of \texttt{tmle\_TSM\_all()} function in the \texttt{tmle3} R package.
We note that because the nuisance function learners implemented by CV-TMLE were not extracted to be included in our selected models, the results might not be fully comparable.  Nevertheless, we include them as a benchmark here.
While our simulations are sufficiently extensive to establish benchmark comparisons with the most widely used estimators in a typical practical setting, more exhaustive simulations in more challenging settings particularly in presence of high-dimensional covariates might also be of interest and could be the topic of future work.}

The relative bias and mean squared error of $\widehat \psi$ of different methods with mixed-minimax as baseline are shown in Tables~1 and 2.
\yifan{The absolute bias and mean squared error of $\widehat \psi$ of different methods are shown in Section~S5 of the Supplementary Material.}
The proposed estimator has the smallest bias across almost all sample sizes, and there is a notable gap between the proposed estimators and most of those estimated by DDML and CV-TMLE. 
This confirms our earlier claim that combined with flexible nonparametric/machine learning methods, our proposed approach has the potential to yield smaller bias than competing methods, without directly estimating the corresponding bias. As can be seen from the mean squared error, the proposed selection does not have the smallest mean squared error, which is not surprising as the proposed selection procedure does not aim to minimize the mean squared error.

We also examined the empirical coverage of 95\% confidence intervals for DDML and the proposed selector at each sample size.
\yifan{The standard error of the proposed estimator is computed by option (i) described in the previous section.
Similarly, the standard errors of DDML are computed using the average of the standard errors of influence functions evaluated at $\widehat \psi_1$ and $\widehat \psi_2$.}
The results are presented in Tables~3 and 4. As can be seen, the coverage of the proposed confidence intervals is slightly lower than nominal coverage, and the proposed confidence intervals are slightly wider due to the selection step, potentially resulting in smaller bias and more accurate confidence intervals.
Furthermore. the random forests provide confidence intervals which have poor coverage in this case. However, as can be seen, the proposed confidence intervals may not be affected by poor performance of one of the candidate learners.

\begin{table}[h]
\begin{center}
{Table~1. Scenario 1: relative bias (relative MSE) with mixed-minimax as baseline $\text{absolute bias}/\text{absolute bias of mixed-minimax}$ ($\text{MSE}/\text{MSE of mixed-minimax}$)}\\
\begin{tabular}{cccccccccc}
       & $n=250$  & $n=500$  & $n=1000$  & $n=2000$  \\
DDML-LASSO    & 3.3 (1.3)  &  6.2 (1.1)  & 27.9 (1.1) &  54.9 (1.3)\\
DDML-RF    & 27.6 (4.8) &    19.6 (2.2) & 54.9 (1.7) & 63.8 (1.4) \\
DDML-GBT  & 1.6 (0.8)   &  1.2 (0.6)  & 13.5 (0.6) &  28.9 (0.6)\\
DDML-SL  & 4.4 (1.1)  &  1.6 (0.7) & 5.2 (0.6) &  22.7 (0.6) \\
CV-TMLE-LASSO    & 2.8 (0.7)  &  5.2 (0.8)  & 25.7 (0.9) &  54.3 (1.2)\\
CV-TMLE-RF    & 8.7 (0.9) &    9.9 (0.9) & 28.7 (0.8) & 34.3 (0.8) \\
CV-TMLE-GBT  & 1.1 (0.6)   &  1.1 (0.5)  & 10.1 (0.5) &  17.7 (0.5)\\
CV-TMLE-SL  & 0.4 (0.5)  &  0.7 (0.5) & 4.5 (0.5) &  5.4 (0.5) \\
\end{tabular}\\
\small{\yifan{DDML-LASSO refers to DDML using logistic regression with $l_1$ regularization for the propensity score, and standard Lasso for the outcome model;
DDML-RF refers to DDML using classification forests for the propensity score, and regression forests for the outcome model;
DDML-GBT refers to DDML using gradient boosting classification tree for the propensity score, and gradient boosting regression tree for the outcome model; 
DDML-SL refers to DDML using super learner for the propensity score, and super learner for the outcome model;
CV-TMLE-LASSO refers to TMLE with a Lasso library for both the propensity score and the outcome model; 
CV-TMLE-RF refers to TMLE with a random forest library for both the propensity score and the outcome model; 
CV-TMLE-GBT refers to TMLE with a gradient boosting tree library for both the propensity score and the outcome model; 
CV-TMLE-SL refers to TMLE with a library consisting of Lasso, random forests, and gradient boosting trees for both the propensity score and the outcome model; 
 Mixed-minimax refers to
the proposed selector.}}
\end{center}
\end{table}

\begin{table}[h]
\begin{center}
{Table~2. Scenario 2: relative bias (relative MSE) with mixed-minimax as baseline $\text{absolute bias}/\text{absolute bias of mixed-minimax}$ ($\text{MSE}/\text{MSE of mixed-minimax}$)}\\
\begin{tabular}{cccccccccc}
       & $n=250$  & $n=500$  & $n=1000$  & $n=2000$  \\
DDML-LASSO    & 1.1 (0.8)  &  1.6 (0.8)  & 0.7 (0.7) &  1.5 (0.7)\\
DDML-RF    &  78.8 (3.7) &    20.9 (2.9) & 40.2 (2.1) & 47.4 (1.4) \\
DDML-GBT  & 3.6 (0.7)   &  0.7 (0.7)  & 1.9 (0.7) &  2.7 (0.6)\\
DDML-SL  & 5.5 (0.7)  &  1.6 (0.7) & 1.3 (0.7) &  5.4 (0.6)\\
CV-TMLE-LASSO    & 0.6 (0.6)  &  0.8 (0.6)  & 0.5 (0.6) &  2.1 (0.6)\\
CV-TMLE-RF    & 11.0 (0.7) &    4.4 (0.7) & 11.1 (0.7) & 18.0 (0.7) \\
CV-TMLE-GBT  & 8.1 (0.6)   &  3.8 (0.6)  & 13.1 (0.6) &  14.8 (0.6)\\
CV-TMLE-SL  & 0.6 (0.5)  &  0.2 (0.6) & 3.5 (0.6) &  3.1 (0.6) \\
\end{tabular}\\
\small{See Table 1 caption for abbreviations.}
\end{center}
\end{table}

\begin{table}[h]
\begin{center}
{Table~3. Scenario 1: error rates in percent and average width of $95\%$ confidence intervals.}\\
\begin{tabular}{cccccccccc}
         & \multicolumn{2}{c}{$n=250$} & \multicolumn{2}{c}{$n=500$} & \multicolumn{2}{c}{$n=1000$} & \multicolumn{2}{c}{$n=2000$}\\
         & WD     & ER     & WD     & ER  & WD  & ER  & WD   & ER   \\
DDML-LASSO    &  1.2   &    7.4   &  0.8  &  7.8 &  0.5 &  6.6 &  0.4 & 10.0 \\
DDML-RF    &  1.6 &  13.4  &  0.9  & 10.4   & 0.6 &  11.0  &  0.4 &  10.6 \\
DDML-GBT & 0.9 & 6.6 & 0.6 &  7.4  & 0.4 & 7.0  & 0.3 & 8.4\\
DDML-SL & 1.0 & 6.2  & 0.6 & 6.0   & 0.4   &  5.2 & 0.3 & 6.6 \\
Mixed-minimax   & 1.1  & 6.2  & 0.7 & 7.0 & 0.5 & 6.4  & 0.4 &  6.8 \\
\end{tabular}\\
\small{WD denotes the average width of the confidence interval; ER denotes the error rate of the confidence interval.}
\end{center}
\end{table}

\begin{table}[h]
\begin{center}
{Table~4. Scenario 2: error rates in percent and average width of $95\%$ confidence intervals.}\\
\begin{tabular}{cccccccccc}
         & \multicolumn{2}{c}{$n=250$} & \multicolumn{2}{c}{$n=500$} & \multicolumn{2}{c}{$n=1000$} & \multicolumn{2}{c}{$n=2000$}\\
         & WD     & ER     & WD     & ER  & WD  & ER  & WD   & ER   \\
DDML-LASSO    &  0.8   &   7.6   &  0.5  &  5.0 &  0.4 &  4.6 &  0.2  & 5.2 \\
DDML-RF    &  1.1 &  19.6  &  0.7 & 16.4  & 0.4 &  14.8  & 0.3 & 12.4 \\
DDML-GBT & 0.7 &  8.4 & 0.5 &  5.6  & 0.3 & 5.2  & 0.2 & 5.4 \\
DDML-SL & 0.7 & 8.8   & 0.5 & 5.4   & 0.3  & 5.0 & 0.2 & 5.2\\
Mixed-minimax   & 0.8  & 9.8  & 0.6 & 6.6 & 0.4 & 7.4  &  0.3 &  8.2 \\
\end{tabular}\\
\small{WD denotes the average width of the confidence interval; ER denotes the error rate of the confidence interval.}
\end{center}
\end{table}

\section{Data Analysis\label{sec:real}}
In this section, similarly to \cite{tan2006,2015biasreduce,tan2020model,tan2020regularized}, we reanalyze data from the Study to Understand Prognoses
and Preferences for Outcomes and Risks of Treatments (SUPPORT) to evaluate the effectiveness of right heart catheterization (RHC) in the intensive care unit of critically ill patients. At the time of the study by \cite{5c6af36c0fb64cfcbb482d75c2bc7ff1}, RHC was thought to lead to better patient outcomes by many physicians. \cite{5c6af36c0fb64cfcbb482d75c2bc7ff1} found that RHC leads to lower survival compared to not performing RHC.

We consider the effect of RHC on 30-day survival. Data are available on 5735 individuals, 2184 treated and 3551 controls. In total, 3817 patients survived and 1918 died within 30 days. To estimate the additive treatment effect  $\psi_0 = E\{Y_1-Y_0\}$, 72 covariates were used to adjust for potential confounding \citep{Hirano2001}. We posited the same 
 candidate models/estimators as in Section~\ref{sec:simu3}. 
The proposed selection procedure was implemented with two-fold cross-validation.
Tuning parameters were selected as in Section~\ref{sec:simu3}.

The proposed selection criterion selected logistic regression with $l_1$ regularization for the propensity score and Lasso for the outcome model.
The estimated causal effect of RHC was $-0.0661$. The result was somewhat smaller than the targeted maximum likelihood estimator and other improved estimators considered by \cite{2015biasreduce}, who did not perform model selection. Specifically,
the targeted maximum likelihood estimator with default super learner \citep{van2011targeted} gave $\widehat \psi_{\text{TMLE-SL}} = - 0.0586$;
  the bias reduced doubly robust estimation with linear and logit link gave $\widehat \psi_{\text{BR},\text{lin}}=-0.0612$ and $\widehat \psi_{\text{BR},\text{logit}}=-0.0610$, respectively;
   the calibrated likelihood estimator \citep{tan2010} gave $\widehat \psi_{\text{TAN}}=-0.0622$.
   Our estimates suggest that previous estimates may still be subject to a small amount of misspecification bias.

Similar to Section~\ref{sec:simu3}, we implemented the 95\% confidence interval, estimated as $(-0.0976,-0.0345)$, which was slightly wider than other improved estimators considered in \cite{2015biasreduce}, e.g., the targeted maximum likelihood estimation with default super learner gave $(-0.0877, -0.0295)$;
the bias reduced doubly robust estimator with linear and logit link gave $(-0.0889,-0.0335)$ and $(-0.0879,-0.0340)$, respectively;
the calibrated likelihood estimator gave $(-0.0924,-0.0319)$. This is not surprising because we consider a richer class of models, potentially resulting in smaller bias and more accurate confidence intervals.

\section*{Acknowledgements}

We thank Rajarshi Mukherjee, James Robins, Andrea Rotnitzky, and Weijie Su for helpful discussions. We thank Karel Vermeulen and Stijn Vansteelandt for providing the dataset.
We thank the three reviewers, associate editor, and editor for many useful comments which led to an improved manuscript. 
Yifan Cui was supported in part by the National Natural Science Foundation of China.

\section*{Supplementary Material}
Supplementary material available at \textit{Biometrika}~online includes all proofs and additional results.

\appendix

\section{Appendix}

\begin{algorithm} 
\SetAlgoLined
\caption{The proposed selective algorithm to estimate average treatment effect} \label{alg:cf}
\KwIn{Dataset $\mathcal O$ and $K \times L$ candidate models;}
\For{$s = 1$ \textbf{to} $S$}
{\ShowLn In the training dataset $\mathcal O^{0s}$: Construct models of $\widehat \pi^s_k(A|X)$ and $\widehat E^s_l(Y|A,X)$  for each $k\in \mathcal K, l \in \mathcal L$\;
\ShowLn In the validation dataset $\mathcal O^{1s}$:  For each pair $(k,l)$, evaluate
$$\widehat \psi_{k,l}^s =  \PP^1_s
\left\{ \frac{\left( -1\right) ^{1-A}}{\widehat \pi^s_{k}\left( A|X\right) }
\left\{
\begin{array}{c}
Y -  \widehat E^s_{l}(Y| A,X)
\end{array}
\right\} +\sum_a (-1)^{1-a}\widehat E^s_{l}(Y|a,X) \right\};$$
  \\}
\ShowLn For each pair $(k_0,l_0)$, average the perturbations over the splits and obtain
$$\widehat {\text{per}}(k,l; k_0,l_0) = \frac{1}{S}\sum_{s=1}^S \left[  (\widehat \psi_{k,l}^s  - \widehat \psi_{k_0,l_0}^s )
\right]^{2};$$
\ShowLn Calculate $$\widehat B_{k_0,l_0} =\max_{l_1,l_2 \in \mathcal L} \widehat{\text{per}}(k_0,l_1; k_0,l_2) + \max_{k_1,k_2 \in \mathcal K} \widehat{\text{per}}(k_1,l_0; k_2,l_0),$$
for each pair $(k_0,l_0)$ \;
\ShowLn Pick $(\widehat k^*,\widehat l^*)=\arg\min_{(k,l)} \widehat B_{k,l}$ as our selected learners, and obtain the estimation of the parameter over the splits  $$\widehat \psi_{\widehat k^*,\widehat l^*}=\frac{1}{S}\sum_{s=1}^S \widehat \psi^s_{\widehat k^*,\widehat l^*};$$\\
\textbf{Return} $(\widehat k^*,\widehat l^*)$ and $\widehat \psi_{\widehat k^*,\widehat l^*}$.
\end{algorithm}

\bibliographystyle{asa}
\bibliography{causal,surv}

\end{document}